\documentclass[preprint, amsmath]{revtex4}
\usepackage{amssymb}
\usepackage{indentfirst}
\usepackage[mathscr]{eucal}
\usepackage{graphicx}

\begin{document}

\title{
``Ping-pong'' electron transfer. \\
I. First reflection of the Loschmidt echo.
      }

\author{
V.N. Likhachev, T.Yu. Astakhova, G.A. Vinogradov${}^*$
       }

\affiliation{
Emanuel Institute of Biochemical Physics, \\
 Russian Academy od Sciences, Moscow, Russian Federation
            }

\begin{abstract}

Quantum dynamics of the electron wave function on one-dimensional
lattice is considered. The lattice consists of $N$ equal sites and
one impurity site. The impurity site differs from other sites by
the on-site electron energy $E$ and the hopping integral $C$. The
wave function is located on the impurity site at $t = 0$. The wave
packet is formed which travels along the lattice, and reflects
from its end. Reflections happen many times (Loschmidt echo)  and
this phenomenon is considered in the second part of the paper.
Analytical expressions for the wave packet front propagation at
different values $E$ and $C$ are derived, and they are in
excellent agreement with the numerical simulation. The obtained
results can help in interpretation of recent experiments on highly
efficient charge transport in synthetic olygonucleotides.

\vspace{2 cm}



\noindent ${}^*$The corresponding author:
\texttt{gvin@deom.chph.ras.ru}.

\end{abstract}

\maketitle


\section{INTRODUCTION}

Charge transfer (CT) is the key event in many processes in living
and inorganic matter. The particular interest has been expressed
after the effective charge transfer in synthetic olygonucleotides
was found \cite{Aug07, Bar11, Gen10, Mal10, Gie02}. There are
possibilities to synthesize the DNA analogues with the bases
sequences up to dozens base pairs. For example, the double strand
consisting of 100 base pairs adenosine--thymine (A-T)${}_{100}$
was synthesized \cite{Sli11}. This synthetic olygonucleotide have
regular structure and very effective with regard to the CT.
Synthetic DNA and polypeptides can be utilized in nanobiology
\cite{Gie06, Gao11}. They are also considered as potential
molecular wires \cite{Mal07}. It is worth noting that the CT can
occur as the one-step coherent process \cite{Gen11}.

Experiments on the charge transfer are usually done when donor $D$
and acceptor $A$ are attached to both sides of the molecular
chain:  $D-(b)_N-A$. The charge is transferred from donor on the
chain and travels along the chain $-(b)_N-$ consisting of $N$
repeating DNA bases (usually adenine). The charge can be trapped
by an acceptor and registered by photophysical response
(fluorescence) or electrochemical reactions.

If an acceptor is absent then the charge is locked on the chain
and can reflect from the end. Many reflections can happen (the
phenomenon of Loschmidt echo, i.e. multiple returning to the
initial state, was extensively studied in classical and quantum
systems \cite{Qua06, Jac02, Cuc03, Pro03, Man06}). V.~Benderskii
with colleagues investigated the analogous quantum dynamics of
vibrational excitations in 1D molecular chain \cite{Ben07, Ben08,
Ben11}. It worth noting that the property to form the moving wave
packet is an attribute of discrete lattices. The diffusive
spreading of the initially localized excitation is usually
observed in continuous models.

The recurrence phenomena to the initial state are well known. This
property is also typical for some dynamical systems. The
Fermi-Pasta-Ulam recurrence is an example \cite{Fer55, Ber05}. The
other example is the interaction of the isolated vibrational mode
with the continuous spectrum \cite{Ovc01}.

The recurrence in the quantum systems was for the first time
considered by R.~Zwanzig \cite{Zwa60} for the discrete
equidistante spectrum when single level is initially populated.
The analysis of the Zwanzig's problem is done in \cite{Ben07}. The
same authors generalized their consideration in the next paper
\cite{Ben08}. The quantum dynamics of the vibrational excitation
on the one-dimensional lattice was considered in \cite{Ben11}. It
was shown that if the excitation is initially localized in the
lattice center, then the energy is not distributed homogeneously
along the lattice but conversely it localizes and reflects from
the lattice ends many times.

In the present paper we consider an analogous problem of the wave
packet propagation on the lattice and multiple reflections from
the lattice ends. One of the goals is an explanation of efficient
charge transport in synthetic olygonucleotides.


\section{Setting up a problem}

We consider a lattice consisting of $N$ identical sites and one
impurity site at the left lattice end. The excitation is initially
confined to the impurity site. An excitation can be electronic or
vibronic. The electronic excitation is considered for
definiteness.

The wave function amplitude on the impurity site is labelled by
$a(t)$, and amplitudes on other sites $b_i(t) \,\, (i=1,2, \ldots,
N)$. The electron can hop on the neighboring site. The electron
energy on all sites except the impurity site is zero, what
corresponds to the choice of the reference point for electron
energy. The hopping integral $C = 1$, what corresponds to the
choice of energy unit. $E$ is the on-site energy of the impurity
site, and $C$ -- the hopping integral between the impurity site
and the nearest site of the lattice. The challenge is to find how
the electronic populations of all sites change in time.

The hamiltonian in the $(N+1) \times (N+1)$ matrix representation
reads:
\begin{equation}
  \label{Ham1}
  H =
\begin{pmatrix}
{\bf E} & {\bf C} & 0 &  \cdots & 0 & 0 & 0 \\
{\bf C} & 0       & 1 &  \cdots & 0 & 0 & 0 \\
0       & 1       & 0 &  \cdots & 0 & 0 & 0 \\
\cdots & \cdots & \cdots  & \cdots & \cdots
& \cdots & \cdots \\
0 & 0 & 0 &  \cdots & 0 & 1 & 0 \\
0 & 0 & 0 &  \cdots & 1 & 0 & 1 \\
0 & 0 & 0 &  \cdots & 0 & 1 & 0 \\
\end{pmatrix}
\end{equation}
with the wave function
\begin{equation}
  \label{WF1}
\Psi(t) =  a(t), b_1(t), b_2(t), \ldots, b_N(t).
\end{equation}

In the second quantization representation the hamiltonian is
\begin{equation}
  \label{Ham2}
   H = E a^+ a^- + \sum_{i=1}^{N-1} b^+_{i+1}b^-_i +
   \sum_{i=2}^{N}b^+_{i-1}b^-_i +
   C \left( a^+b_1^- + b_1^+ a^- \right)\,.
\end{equation}

The dimensionless  ($\hbar = 1$) Schr\"odinger
equation for the wave function has the form
\begin{equation}
  \label{Sch1}
  \left\{
  \begin{split}
  i \dot a = & \, E a + C b_1 \\
  i \dot b_1 = & \, b_2 + C a \\
  i \dot b_2 = & \, b_1 + b_3 \\
 \cdots \,\,\,\, &  \, \cdots \cdots \cdots   \\
  i \dot b_N = & \, b_{N-1} \\
  \end{split}
  \right.
\end{equation}
with the initial conditions $\left. a(t) \right|_{t=0} = 1$ and
$\left. b_i(t) \right|_{t=0} = 0 \,\, (i = 1,2, \ldots, N)$. The
formulated problem mimics the experiments on the charge transfer
in synthetic DNA \cite{Bar11}.

\begin{figure}
\begin{center}
  \includegraphics[width=100mm,angle=0]{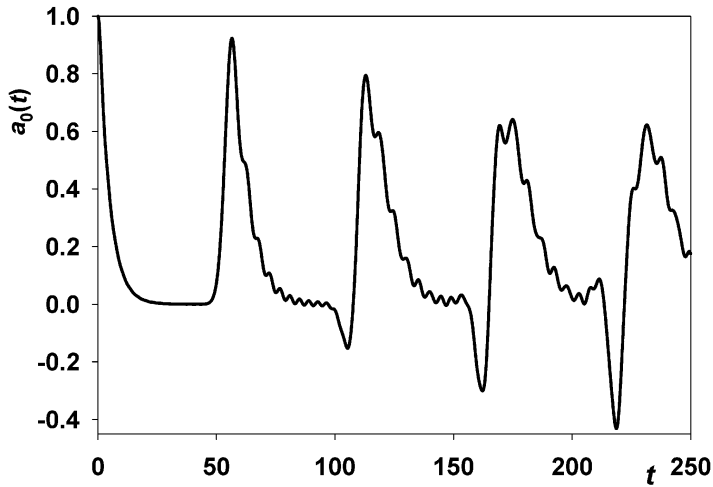}
 \caption{
The dependence of the wave function amplitude $a(t)$ on the
impurity site vs. time. The wave function returns many times (four
times in the figure) to the impurity site with slightly decreasing
amplitudes. Parameters: $E=0$, $C= \sqrt{2}$, $N = 50$.
         }
  \label{Fig_00_}
  \end{center}
\end{figure}

If equations \eqref{Sch1} are integrated numerically then the
phenomenon of repeated reflections from the lattice ends is
observed, i.e. Loschmidt echo or the electronic ``ping-pong'' (the
electronic ping-pong was found experimentally \cite{Eli08}). The
result is shown in Fig.~\ref{Fig_00_}.

The goal of the present paper is the consideration of the wave
packet propagation and its first reflection from the lattice end.
This time interval corresponds to the time range $0 \leq t < 100$
in Fig.~\ref{Fig_00_}. Next part of the paper considers the
multiple reflections of the wave packet.

The quantum dynamical problem of the wave packet evolution can be
treated as the interaction of the impurity site with the
reservoir. And it is solved using the expansion in terms of the
eigenfunctions of the reservoir.

The lattice without the impurity site (reservoir) is described by
the $N \times N$ tridiagonal matrix with zero leading diagonal and
the unity values on the secondary diagonals. Eigenvalues
$\varepsilon(k)$ and eigenfunctions $q_i(k)$ of this matrix are
well known:
\begin{equation}
  \label{Eigen1}
  \varepsilon(k) = 2 \, \cos \left( \dfrac{\pi k}{N+1} \right);
  \quad
  q_i(k) = \sqrt{\dfrac{2}{N+1}} \sin \left( \dfrac{\pi k}{N+1}i
  \right).
\end{equation}

It is more convenient to consider the problem in terms of
amplitudes of modes $b(k)$ instead of amplitudes on the lattice
sites $b_i$. These values are related by the orthogonal
relationship:
\begin{equation}
  \label{Ortho}
 b(k,t) \equiv \sum_{i=1}^{N} q_i(k) \, b_i(t);
  \quad
  b_i(t) \equiv \sum_{k=1}^N q_i(k) \, b(k,t).
\end{equation}
Then the system \eqref{Sch1} can be written in the form:
\begin{equation}
  \label{Sch2}
  \left\{
  \begin{split}
  i \dot a(t) = & \, E a(t) + C \sum_{k=1}^N q_1(k) \, b(k,t) \\
i \dot b(k,t) = & \, \varepsilon (k) \, b(k) + C a(t) \, q_1(k) \\%
  \end{split}
  \right.
\end{equation}

Amplitudes $b(k,t)$ in \eqref{Sch2} are expressed through the
amplitude $a(t)$:
\begin{equation}
  \label{Tr_01}
 b(k,t) = q_1(k) \exp[-i \, \varepsilon(k) \, t] \,
 \int_0^t a(\tau) \, \exp[i \, \varepsilon(k) \, \tau] \, {\rm d}
 \tau \,.
\end{equation}
Substituting this expression into equation \eqref{Sch2} for
$a(t)$, one can get the integro-differential equation for the
amplitude $a(t)$ on the impurity site:

\begin{equation}
  \label{Tr_02}
    {\dot a(t)} = - i \, E \, a(t) - C^2
  \int_0^t B^N(t-t') \, a(t'){\rm d} \, t',
\end{equation}
where the kernel is given by the following sum:
\begin{equation}
  \label{Tr_03}
 B^N(t) = \dfrac{2}{N+1} \sum_{k=1}^N
 \sin^2 \left( \dfrac{\pi k}{N+1} \right)
\exp \left[ - 2 \, i \, \cos \left( \dfrac{\pi k}{N+1}  \right) t \right].
\end{equation}

Our primary goal is the solution of \eqref{Tr_02} for the
amplitude $a(t)$ on the impurity site.


\section{The decay of the initial state $a(t=0) = 1$
in the semi-infinite lattice}

Initially we consider the decay kinetics of the initial state
$\left. a(t) \right|_{t=0} = 1$ in the semi-infinite lattice ($N
\to \infty$). The amplitude on the impurity site is labelled by
$a_0(t)$ in this case. In the expression \eqref{Tr_03} for the
kernel $B^N(t)$ we use the limit $N \to \infty$. The limiting
value (at $N \to \infty$) of the expression for $B^N(t)$ is
labelled by $B_0(t)$.

The sum \eqref{Tr_03} in the limit $N \to \infty$ is modified to
the integral \cite{Gra07}:
\begin{equation}
  \label{B0_1}
 B_0(t) \equiv \lim B^N(t) \left. \right|_{N \to \infty} = J_0(2t) + J_2(2t).
\end{equation}
And in this limit we have the following equation for the amplitude
$a_0(t)$:
\begin{equation}
  \label{a0_1}
\dot a_0(t) = - i E a_0(t) - C^2 \int_0^t B_0(t-t') \, a_0(t') \,
{\rm d}t'.
\end{equation}
This equation can be solved using the Laplace transformation:
\begin{equation}
  \label{Lapl_1}
a_0(p) = \dfrac{1}{p + i E + C^2 \, B_0(p)}\,,
\end{equation}
where the Laplace transform $B_0(p)$ of the function $B_0(t)$
\cite{Gra07}:
\begin{equation}
  \label{Lapl_2}
 B_0(p) = \dfrac12 \, \left( \sqrt{p^2 + 4} - p  \right).
\end{equation}

The amplitude $a_0(t)$  can be obtained by the inverse Laplace
transformation:
\begin{equation}
  \label{Lapl_3}
 a_0(t) = \int\limits_{d - i \infty}^{d + i\infty}
 \dfrac{\exp(p\,t)}{p + iE + C^2 \dfrac12
 \left( \sqrt{p^2 + 4} - p \right)} \, {\rm d}p,
 \quad d > 0.
\end{equation}

The obtained form of the expression for $a_0(t)$ is very
inconvenient for the analysis and numerical computations (integral
with the upper infinite limit converges very slowly). The
convenient form this integral obtains if the integration contour
is closed around the cut $[-2i,+2i]$, and the square root
$\sqrt{p^2 + 4}$ should be rewritten in the form: $\sqrt{p^2 + 4}
= -i \, \sqrt{ip+ 2} \, \sqrt{ip - 2}$. The integration contour
can be closed if amplitude $a_0(p)$ has no poles. Assuming that
poles are absent, close the integration contour and summing the
integrals along both banks of the cut, one gets the following
expression for amplitude $a_0(t)$:
\begin{equation}
  \label{a0_2}
  a_0(t) = \dfrac{1}{\pi} \int_{-2}^{+2} {\rm d} \omega
  \exp(i \omega t) \, {\rm Im}
  \left[ \omega + E - \dfrac{C^2}{2}
  \left( \omega + i \sqrt{4 - \omega^2}  \right)
  \right]^{-1}.
\end{equation}

Few special cases can be noted, when the integral \eqref{a0_2} is
calculated explicitly and the amplitude $a_0(t)$ is expressed
through the Bessel functions \cite{Gra07}:
\begin{equation}
  \label{a0_3}
  \begin{split}
a_0(t) = & J_0(2t);\qquad \qquad \qquad (E = 0, \,\, C = \sqrt{2})
\\
  a_0(t) = & J_0(2t) + J_2(2t); \qquad (E = 0, \,\, C = 1)  \\
 a_0(t) = & J_0(2t) - i J_1(2t); \qquad (E = 1, \,\, C = 1)  \\
  \end{split}
\end{equation}

An integro-differential equation for the amplitude  $a_0(t)$ is
derived in Appendix A, and its solution is obtained as a series in
terms of the Bessel functions. These series have simple form in
few particular cases when $E=0$ or  $C=1$.

\begin{figure}
  \begin{center}
  \includegraphics[width=100mm,angle=0]{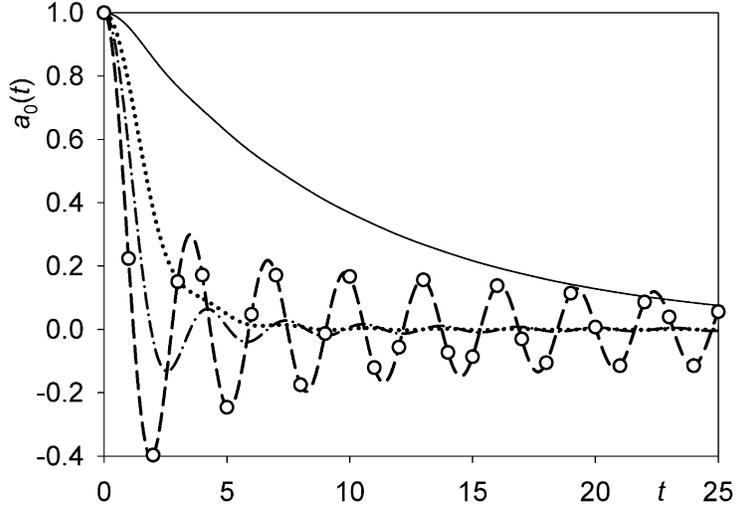}
 \caption{
The dependence of the amplitude $a_0(t)$, calculated according to
\eqref{a0_2}, vs. time at $E=0$. Solid line: $C^2 = 0.1$; dotted
line: $C^2 = 0.5$; dash-dotted line: $C^2 = 1$; dashed line:  $C^2
= 2$. Empty circles -- accurate result for $C^2 = 2$, when
$a_0(t)=J_0(2t)$.
         }
  \label{Fig_01_}
  \end{center}
\end{figure}

Fig.~\ref{Fig_01_} shows the dependence of the amplitude $a_0(t)$
vs. time at $E=0$ and at different values of the hopping integral
$C^2$. The comparison with the accurate result is also shown
(``accurate'' result $\equiv$ numerical integration of equations
\eqref{Sch1}).

The dependence of the amplitude vs. time in \eqref{a0_2} can be
estimated by the Fermi's golden rule, when the perturbation
approximation by the small parameter $C^2$ is used, and when an
exponential decay is valid:
\begin{equation}
  \label{Fer}
  a_0(t) \approx \exp
  \left( - C^2  \sqrt{4 - E^2} \, t \right).
\end{equation}
More accurate expression for the amplitude $a_0(t)$ at small
values of parameter $C$ is given in Appendix B.

\begin{figure}
\begin{center}
  \includegraphics[width=70mm,angle=0]{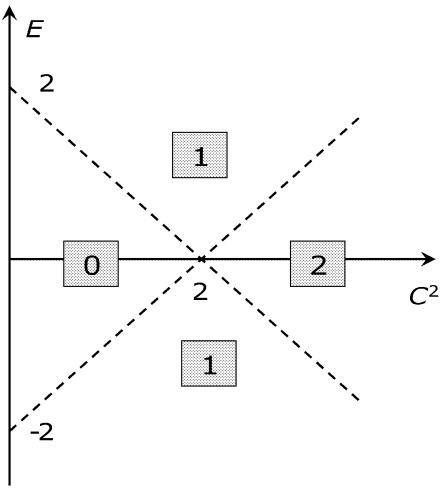}
 \caption{
Number of poles of the integrand of function \eqref{a0_2} in
regions divided by dashed lines.
         }
   \label{Fig_02_}
  \end{center}
\end{figure}

Now we consider the case when amplitude $a_0(p)$ has poles. An
analysis shows that one pole exists in the region $E>2-C^2$, and
two poles -- when $E<C^2-2$. There are no poles in the region $|E|
< 2 - C^2$ ($C^2<2$). Fig.~\ref{Fig_02_} shows number of poles vs.
values of parameters $E$ and $C$.

As an example we consider the case of one pole in the region
$E>2-C^2$ ($C^2<2$). The pole is located at the point $p = -i
\varepsilon$ ($\varepsilon >2$). The relation between parameters
$E$ and $C$ follows from \eqref{Lapl_3}:
\begin{equation}
  \label{Rel_1}
  E = \varepsilon  \left( 1 - \dfrac{C^2}{2} \right) +
  \dfrac{C^2}{2} \, \sqrt{\varepsilon^2 -4}.
\end{equation}
The contribution from the pole $\Delta a_0$ is equal to
\begin{equation}
  \label{Rel_2}
  \Delta a_0(t) = \dfrac{\exp(-i \varepsilon t)}
  {1 + \dfrac{C^2}{2} \left( \dfrac{\varepsilon}
  {\sqrt{\varepsilon^2 -4}} -1  \right)}.
\end{equation}

\begin{figure}
\begin{center}
  \includegraphics[width=100mm,angle=0]{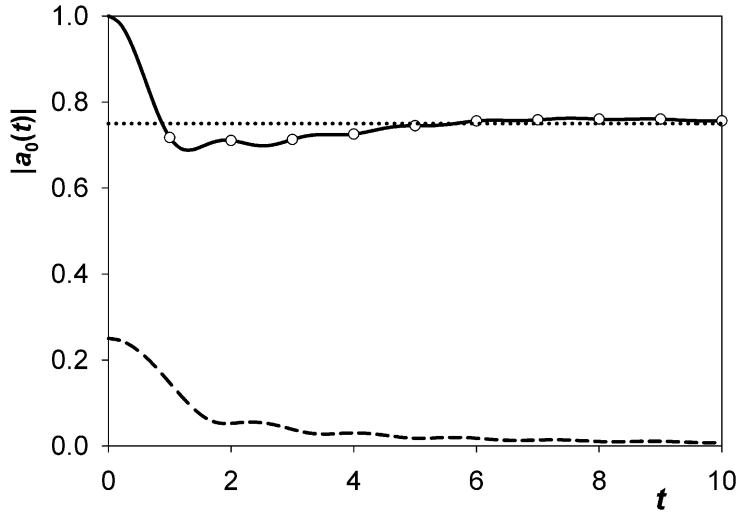}
  \caption{
The dependence of amplitude $a_0(t)$ vs. time in the presence of
one pole. Dashed line -- contribution to amplitude from the
integral summand \eqref{a0_2}; dots -- modulus of the pole summand
(equal to 0.75) \eqref{Rel_2}; empty circles -- numerical result.
Parameters: $E = 2, \,\, C = 1 \,\, (\varepsilon =2.5)$.
          }
    \label{Fig_03_}
  \end{center}
\end{figure}

The overall amplitude is the sum of the main term \eqref{a0_2} and
the pole contribution \eqref{Rel_2}. The dependence of amplitude
vs. time when one pole exists, is shown in Fig.~\ref{Fig_03_}. In
this case the amplitude on the impurity site does not decrease to
zero and $\left. a_0(t) \right|_{t \to \infty} = 0{.}75$.

In Appendix C it is shown that if $|E| > 2 - C^2$  ($C^2<2$), then
there exists the bounded state localized at the lattice end. The
value $\varepsilon$ (see \eqref{Rel_1}) is the energy of this
localized state and its contribution to the amplitude $a_0(t)$
coincides with the contribution from the pole summand. Note also
that the integral term \eqref{a0_2} is the contribution to the
amplitude from the continuous spectrum which lies in the range
$[-2,\, 2]$. The fraction of the initial state is trapped by this
localized state and the smaller part of the initial state goes out
to the reservoir. The effect of returning to the initial state
also decreases. This case (existence of a pole) seams to be less
interesting and we consider only an absence of the localized state
when the integral \eqref{a0_2} is the accurate result for the
amplitude $a_0(t)$.

Such a detailed analysis of the amplitude $a_0(t)$ decay in the
infinite lattice is done by two reasons. First,  the decay
kinetics in the infinite lattice coincides with the kinetic in the
finite lattice consisting of $N$ sites in the time range $0 \leq t
\lesssim N$. Second, as will be shown in the second part of the
paper, the overall amplitude in the finite lattice can be
represented as a sum of partial amplitudes, first of which is just
the amplitude $a_0(t)$.


\section{Wave packet propagation on the lattice}

\begin{figure}
\begin{center}
  \includegraphics[width=100mm,angle=0]{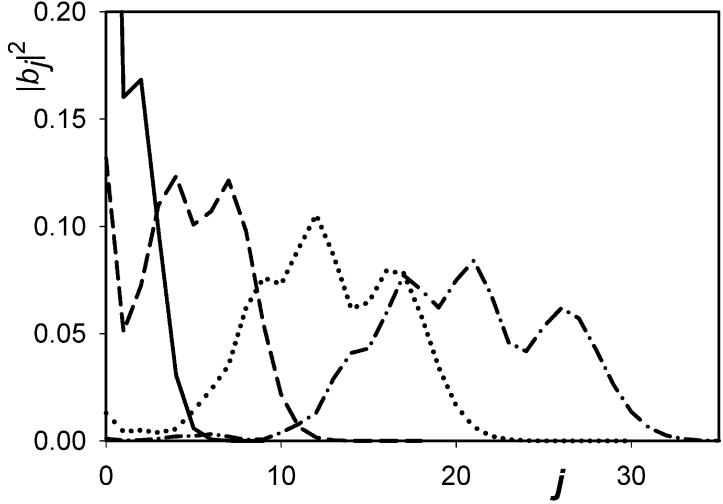}
 \caption{
Spatiotemporal evolution of the wave packet. Solid line: $t = 2$;
dashed line: $t = 5$; dotted line: $t = 10$; dash-dotted line: $t
= 15$. Site $j=0$ corresponds to $|a_0(t)|^2$. Parameters: $E=1$,
$C^2=0.5$. $N = 100$.
         }
    \label{Fig_04_}
 \end{center}
\end{figure}

The dynamics of the wave function on the lattice, i.e.
spatiotemporal evolution of amplitudes $b_j(t)$, is considered in
this section. Initially the expected spreading of the wave packet
is observed. But then amplitudes on the sites with larger numbers
increase. And when the amplitude $a(t)$ becomes small, the wave
function forms the well defined wave packet with the sharp forward
front. This behavior is shown in Fig.~\ref{Fig_04_}.

An expression for amplitudes $b_j(t)$ is obtained by the
substitution expression \eqref{Tr_01} for the mode amplitudes
$b(k,t)$ into expression \eqref{Ortho}. Then one gets the
following equation
\begin{equation}
  \label{b_1}
 b_j(t) = - \dfrac{i 2 C}{N+1}
 \sum_{k=1}^N \sin \left( \dfrac{\pi k}{N+1} \right)
 \sin \left( \dfrac{\pi k}{N+1} j \right)
 \exp[- i \varepsilon(k) t]
 \int_0^t \exp[- i \varepsilon(k) \tau] a(\tau) {\rm d} \tau .
\end{equation}

Initially we consider the case of semi-infinite lattice ($N \to
\infty$). Changing the summation in \eqref{b_1} by integration and
replacing $a(t)$ by $a_0(t)$, the following expression can be
obtained:
\begin{equation}
  \label{b_2}
 b_j(t) = - \dfrac{i 2 C}{\pi}
\int_0^{\pi} {\rm d} k \sin(k) \sin(kj) \exp[- i \varepsilon(k) t] \int_0^t \exp[- i \varepsilon(k) \tau] a(\tau) {\rm d} \tau \,,
\end{equation}
where $\varepsilon(k) = 2 \cos(k)$.

Consider the case when time $t$ is large and amplitude $a_0(t)$ is
very small, $|a_0(t)| \ll 1$. Then the upper limit in integration
over time in \eqref{b_2} is infinity. The obtained integral is the
Laplace transformation $a_0(p)$ taken at $p= - i \varepsilon(k)$.
As a result we have:
\begin{equation}
  \label{b_3}
 b_j(t) = - \dfrac{ 2 C}{\pi}
 \int_0^{\pi} {\rm d} k
 \dfrac{\sin(k) \sin(kj) \exp[- 2 i t \cos(k)]}
 {E - 2 \cos(k) + C^2 \exp(-ik)}, \qquad
 |a_0(t)| \ll 1 \,.
\end{equation}

\begin{figure}
\begin{center}
  \includegraphics[width = 100mm, angle=0]{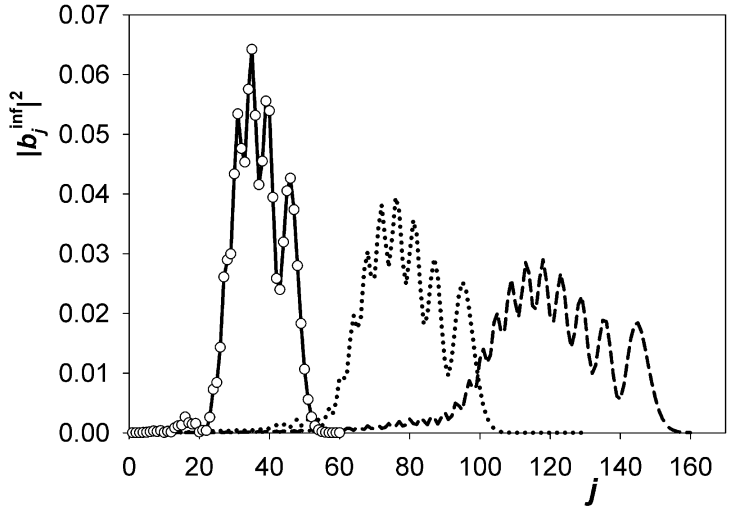}
 \caption{
Snapshots of the impulse in semi-infinite lattice at $t=25$ (solid
line); $t=50$ (dotted line); $t=75$ (dashed line). Calculations
according to \eqref{b_4}. Empty circles -- numerical integration
of \eqref{Sch1}. The coincidence of the numerical and analytical
results is also excellent for $t = 50$ and $t = 75$ (mean square
error (MSE) $\lesssim 10^{-4}$).
         }
 \label{Fig_05_}
  \end{center}
\end{figure}

Expression \eqref{b_3} can formally be generalized on the negative
values of index $j$. Aiming this in mind, the multiplier
$\sin(kj)$ is represented as the difference of two exponents. Then
expression \eqref{b_3} describes the superposition of two wave
packets freely travelling to the left and to the right in the
infinite (to both sides) lattice. Every wave packet is normalized
to unity. Consider separately the wave packet propagating to the
right. Label this wave packet by $b_j^{\rm inf}$:
\begin{equation}
  \label{b_4}
 b_j^{\rm inf}(t) = - \dfrac{ i C}{\pi}
 \int_0^{\pi} {\rm d} k
 \dfrac{\sin(k) \exp\left\{-  i [kj + 2\cos(k)\,t]\right\}}
 {E - 2 \cos(k) + C^2 \exp(-ik)}, \qquad
 -\infty < j < +\infty \,.
\end{equation}
The considered impulse $b_j$ (in the region $j > 0$) is the
following difference:
\begin{equation}
  \label{b_5}
 b_j(t) = b_j^{\rm inf}(t) - b_{-j}^{\rm inf}(t).
\end{equation}
The addend $b_{-j}^{\rm inf}(t)$ is the ``tail'' of the impulse
$b_j(t)$, and it is small at large times, e.g. $ \sim 1/t$. Then
it can be assumed that $b_j^{\rm inf}(t)$ is a good approximation
for $b_j(t)$. The snapshots of the impulse at different time
points are shown in Fig.~\ref{Fig_05_}. One can see that the
constrain by the impulse $b_j^{\rm inf}(t)$ is a very good
approximation.

Expression $b_j^{\rm inf}(t)$ (see \eqref{b_4}) is a reliable
approximation for the propagating impulse on the time interval  $t
\lesssim N/2$ for the finite but comparatively long lattices.
Moreover, as will be demonstrated below, this expression allows to
describe the impulse reflection from the lattice end on the time
interval $N/2 \lesssim t  \lesssim N$.

Lets analyze the expression \eqref{b_4} at large times. As the
integrand is the fast oscillating function, the stationary phase
method can be applied. If $j < t/2$, then two stationary points
exist which are determined by the equality $\sin(k)=j/2t$. Then
the following expression can be obtained:
\begin{equation}
  \label{b_6}
 \widetilde b_j(t) \approx
 \dfrac{-iC}{\sqrt{\pi t \cos(k_0)}}
 \left[
   f(\pi - k_0) \exp(i \pi/4) + f(\pi - k)\exp(- i \pi/4)
 \right], \quad k_0 \equiv \arcsin(j/2t).
\end{equation}
Here $f(k)$ is the integrand in expression \eqref{b_4}.

\begin{figure}
\begin{center}
  \includegraphics[width=100mm,angle=0]{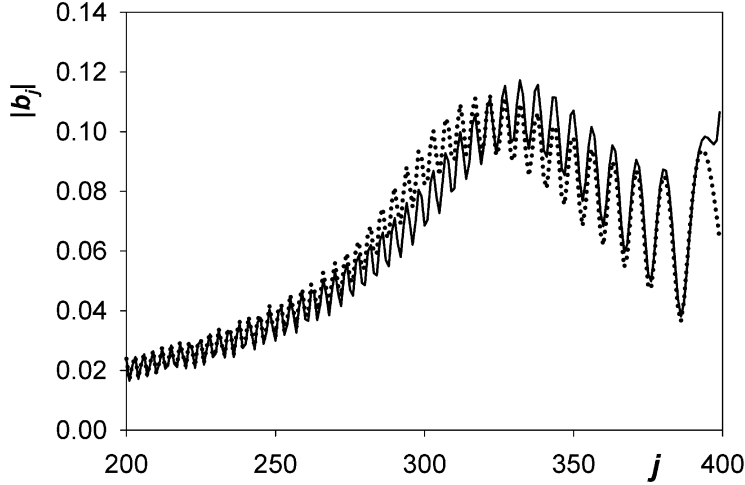}
 \caption{
The impulse front at $t = 200$ $(E=1, \, C^2=2, \, N = 500)$.
Dotted line -- numerical result, solid line -- the asymptotic
expansion\eqref{b_8}.
         }
    \label{Fig_06_}
  \end{center}
\end{figure}

Fig.~\ref{Fig_06_} compares the expression \eqref{b_6}, obtained
by the stationary phase method, with the accurate answer.

If $j \approx 2t$, then both stationary points lie close to the
point $k = \pi/2$ and the stationary phase method gives false
results (if both points coincide then the answer diverges). Lets
consider this region $(j \approx 2t)$ in more details. In this
case the expression for amplitudes \eqref{b_4} can be simplified:
\begin{equation}
  \label{b_7}
 \widetilde b_j(t) \approx
 - \dfrac{i C \exp(- 0.5 \pi j)}{\pi (2 - C^2)}
 \int_{-\infty}^{\infty} {\rm d}x
 \dfrac{\exp \left[
i \left( \dfrac{j - 2t}{\sqrt[3]{t}} x + \dfrac13 x^3
\right)\right] }
 {z - x}, \quad z(t) \equiv \dfrac{(E - i C^2) \sqrt[3]{t}}{2 -
 C^2}.
\end{equation}
If $|z(t)| \gg 1$ (what is valid for $t \gg 1$), then the
asymptotic expansion of \eqref{b_6} can be rewritten as a series
in Airy functions:
\begin{equation}
  \label{b_8}
 \widetilde b_j(t) \approx
 - \dfrac{2 i C }{z(t) (2 - C^2)}
 \left[
{\rm Ai}(\widetilde j) + \dfrac{1}{i z(t)} {\rm Ai}'(\widetilde j)
-
   \dfrac{1}{z^2(t)} {\rm Ai}''(\widetilde j) + \ldots
 \right], \quad \widetilde j = \dfrac{j - 2t}{\sqrt[3]{t}}.
\end{equation}
Thereafter an approximate expression for the impulse front can be
written. Using the asymptotics of the Airy function, one can
notice that the dominant term, describing the impulse front, has
the form:
\begin{equation}
  \label{b_9}
 \widetilde b_j(t) \approx
 - \dfrac{1}{z(t)}
 \exp \left[
  -\dfrac{2}{3} \left( \dfrac{j - 2t}{\sqrt[3]{t}} \right)^{2/3}
 \right].
\end{equation}

\begin{figure}
\begin{center}
  \includegraphics[width=100mm,angle=0]{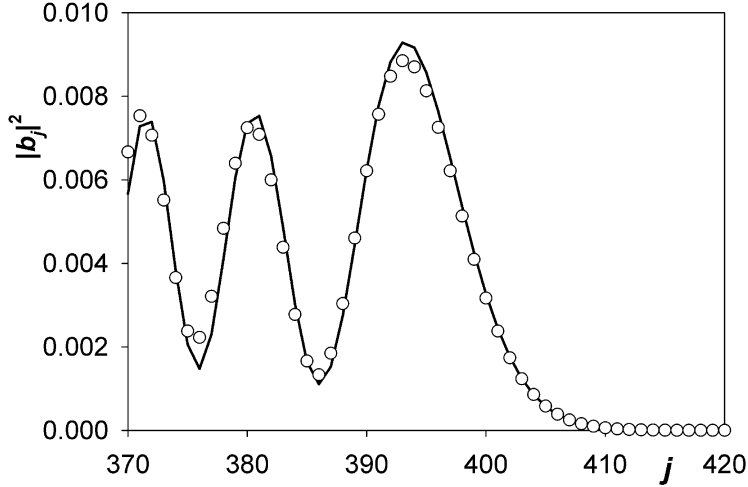}
 \caption{
The impulse front at $t = 200$ $(E=1, C^2=2)$. Empty circles --
numerical result, solid line -- asymptotic expansion \eqref{b_8}.
MSE $\lesssim 10^{-4}$.
         }
  \label{Fig_07_}
  \end{center}
\end{figure}

Fig.~\ref{Fig_07_} shows the comparison of the asymptotic
expansion \eqref{b_8} with the numerical answer.

Expression \eqref{b_9} allows to describe qualitatively the form
of the impulse front. The dependence on the lattice number $j$ is
defined by the argument $(j - 2t)/\sqrt[3]{t}$ (see~\eqref{b_7}).
From \eqref{b_8} one can see that the impulse front slowly
spreads: its width increases $\propto\sqrt[3]{t}$ and amplitude
decreases $\propto 1/\sqrt[3]{t}$. The front velocity is 2 --
maximal possible group velocity. The front has the sharp profile.
Thus, if the finite lattice is considered, then until the front
achieves the lattice end, its propagation is the same as in
infinite lattice. Front of the impulse achieves the lattice end at
$t \approx N/2$. After that the impulse reflects from the lattice
end and the front moves back to the impurity site.

Now we consider the impulse reflection. Lets assume that the
lattice is long enough. Then amplitude $a(t)$ is negligible and
impulse is totally formed. The impulse front moving with the
velocity $v=2$, reaches the lattice end at $t \approx N/2$. After
reflection it moves with the same velocity $v=2$ in the opposite
direction. And at $t \approx N$ the impulse reaches the impurity
site on the left end. Amplitude $a(t)$ starts to increase. We
consider the impulse evolution at time range $t<N$, when amplitude
$a(t)$ is small and when it coincides with the amplitude $a_0(t)$
--- amplitude in the infinite lattice.


\section{Impulse reflection}

To derive an expression for the reflected impulse we come back to
the expression \eqref{b_1} for amplitudes $b_j(t)$. We assume that
the time is large such that amplitude $a(t)$ is small. Then the
upper limit in integration over time is infinity and the
substitution $a(t) \to a_0(t)$ is made. As earlier, the obtained
integral can be represented as the Laplace transformation of
$a_0(p)$ at $p = - i \varepsilon(k)$. Then the following
expression can be deduced:
\begin{equation}
  \label{b_10}
  b_j(t) = - \dfrac{2C}{N+1} \sum_{k=1}^N
  \sin(\widetilde k) \sin(\widetilde k j)
  \exp[- 2 i t \cos(\widetilde k)]
  \left( E - 2 \cos(\widetilde k) +
  C^2 \exp(- i \widetilde k) \right)^{-1},
\end{equation}
where $\widetilde k = \pi k/(N+1)$.

This expression can be modified using the Poisson summation
formula:
\begin{equation}
  \label{Poi}
\sum_{k = -\infty}^{k = \infty} f(k) = \sum_{m = -\infty}^{m =
\infty}
  \int\limits_{-\infty}^{\infty}
  f(x) \, \exp(- 2 \pi i m x) \, {\rm d}x
\end{equation}
and the function $f(k)$ differs from 0 on the interval $[0, N+1]$.
After substitution of variables $\pi x/(N+1) \equiv y$, the
following series can be obtained:
\begin{equation}
  \label{b_11}
  b_j(t) = - \dfrac{2C}{\pi} \sum_{m=-\infty}^{m=\infty}
  \int\limits_{0}^{\pi}
   \dfrac{\sin y \sin(jy)
   \exp \left\{ -i [2t \cos y + 2 m y (N+1)] \right\} }
   {E - 2 \cos y +C^2 \exp(-iy)} {\rm d}y.
\end{equation}
This series can be represented as a sum of two series, expressed
through the amplitude $b_j^{\rm inf}$ --- the amplitude in the
infinite lattice at large time (see \eqref{b_4}):
\begin{equation}
  \label{b_12}
b_j(t) = \sum_{m=-\infty}^{m=\infty} b_{j + 2m(N+1)}^{\rm inf}(t)
-
  \sum_{m=-\infty}^{m=\infty} b_{-j + 2m(N+1)}^{\rm inf}(t)
\end{equation}

\begin{figure}
\begin{center}
  \includegraphics[width=100mm,angle=0]{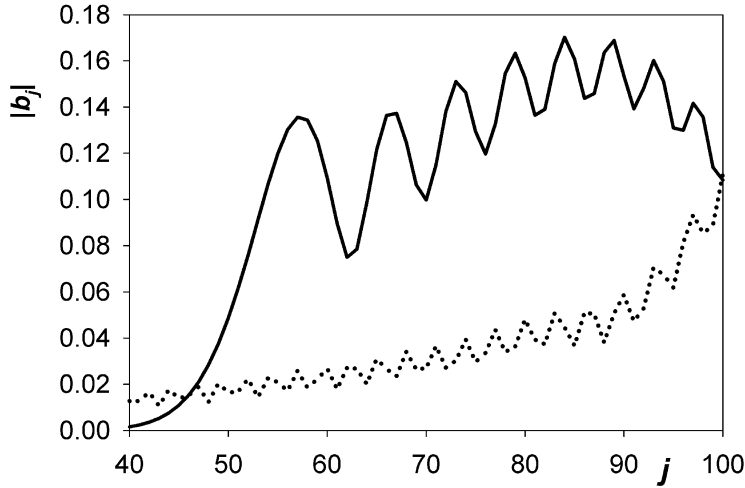}
 \caption{
Reflection of the impulse from the lattice end ($N=100$). Time
$t=75$. Parameters: $E=1, \,\, C^2=0.5$. Computation according to
\eqref{b_6}. Solid line -- reflected impulse, dotted line -- tail
of the incident impulse. At $t = 75$ the impulse front, moving
with $v = 2$, passed $\approx 150$ lattice sites: 100 sites before
reflection and 50 sites after reflection. An excellent agreement
with numerical simulation is observed (MSE $< 10^{-4}$, numerical
data are not shown).
  }
   \label{Fig_08_}
  \end{center}
\end{figure}

For the separating terms, which are significant in the considered
time range $(1 \ll t<N)$, the expression \eqref{b_4} should be
analyzed and the terms, where the phase has the stationary point,
should be found. From the first sum the single term with $m=0$ is
left which describes the incident impulse  $b_j^{\rm inf}$. From
the second sum -- the term with $m=1$, i.e. $b_{-j+2(N+1)}^{\rm
inf}$ describing the reflected impulse. Thus, if $t<N$ then the
impulse can be represented as the sum of two terms:
\begin{equation}
  \label{b_13}
  b_j(t) =  b_{j}^{\rm inf}(t) - b_{-j + 2(N+1)}^{\rm inf}(t).
\end{equation}
Fig.~\ref{Fig_08_} shows the impulse reflection calculated
according to \eqref{b_13}.

At $t \approx N$ the impulse front returns to the lattice
beginning and starts to interact with the impurity site. Amplitude
$a(t)$ increases. Our approximation concerning the smallness of
$a(t)$ becomes invalid. The system returns to the initial state.
This process will be analyzed in the next paper.


\section{Conclusions}

The electronic impulse propagates starting from the impurity site
at the lattice end. If the electronic excitation is entirely
located on the impurity site then, depending on the impurity site
parameters, few scenarios of the wave function dynamics are
realized. One, most common case, is the formation of impulse and
its propagation along the lattice. The impulse moves with the
maximal group velocity $v = 2$ and has the steep forward front.
After reaching the opposite lattice end, the impulse reflects and
moves in the opposite direction. This process repeats many times.
The other scenario is realized, if the parameters of the impurity
site are such that the bounded state is formed. Then the wave
function is partially trapped by this bounded state.

The analytical expressions describing the impulse dynamics
coincide with the accurate numerical simulation.

The problem of the electronic excitation travelling along the
lattice, being formulated in a rather simple form, can be relevant
to recent experiments on the charge transfer in synthetic
olygonucleotides and polypeptides.

At one time it seemed that the polaron mechanism can adequately
explain the charge transport in biomacromolecules. But there
exists at least to weak points in this approach. First, -- polaron
formation. After a charge is transferred to the chain it should be
more or less localized for some time to allow the polaron
formation. Otherwise the wave function spreads very fast along the
lattice because of very small hopping time. Second problem, -- how
the polaron can get an initial momentum necessary for the coherent
movement. The discussed mechanism of quantum dynamical charge
transfer allows to exclude this difficulties in explaining the
high efficient CT over long distances: the electronic wave packet,
responsible for the charge transport, spontaneously forms and
moves.


\appendix

\section{}

An expression for amplitude (see \eqref{Lapl_3}) can be rewritten
in the following form:
\begin{equation}
  \label{a_05}
  a_0(t) = \dfrac{C^2}{2 \pi} \int\limits_{-2}^2 {\rm d} \omega
  \dfrac{ \sqrt{4 - \omega^2} \exp(i \omega t)}
  {(\omega + E) \, (\omega + E - \omega C^2) + C^4}.
\end{equation}
This expression can be rearranged to the differential equation by
time. It can be done if the denominator of the integrand is
represented as the differential operator with the replacement
$\omega \to -i{\rm d}/{\rm d}t$. Then the action of this
differential operator on both sides of equation \eqref{a_05}
eliminates the denominator. The resulting integral is expressed
through the Bessel functions. As a result the following
differential equation is obtained:
\begin{equation}
  \label{DifEq1}
  (1 - C^2) \, \ddot a_0 + i E (2 - C^2) \, \dot a_0 -
  (C^4 + E^2) \, a_0 = - C^2 \left[ J_0(2t) + J_2(2t) \right].
\end{equation}
Hence the inhomogeneous linear differential equation of the second
order is derived. Initial conditions for this equation are:
$a_0(t=0) = 1$ and ${\rm d}a_0/{\rm d}t|_{t=0} = - iE$ (the second
condition follows from equation \eqref{a0_1}).

The solution of this equation can be formally written as the
integral over time. But this integral is difficult to evaluate
analytically and numerical calculation is more complex compared to
the original expression \eqref{Lapl_1}.

Nevertheless we demonstrate that the solution can be represented
through the Bessel functions. And there exist few particular cases
when the solution can be written in the rather compact form.

If $C=1$  then \eqref{DifEq1} is the first order equation. Its
solution can be expanded into a series in Bessel functions:
\begin{equation}
  \label{a_06}
  a_0(t) = J_0(2t) - i E J_1(2t) + (1 - E^{-2}) \,
  \sum_{k=2}^{\infty} (-iE)^k \, J_k(2t).
\end{equation}

In the case when $E=0$ the equation \eqref{DifEq1} has no first
derivative and its solution is a series in even Bessel functions:
\begin{equation}
  \label{a_07}
  a_0(t) = J_0(2t) + (2 - C^2)
  \sum_{k=1}^{\infty} (1 - C^2)^{k-1}\, J_{2k}(2t).
\end{equation}

In the general case the solution of \eqref{DifEq1} can be written
as the following series:
\begin{equation}
  \label{a_08}
  a_0(t) = J_0(2t) - i E J_1(2t) - \dfrac{C^2}{2a(1 - C^2)}
  \sum_{k=2}^{\infty} (b_1 x_1^k - b_2 x_2^k) \, J_k(2t)\,,
\end{equation}
where
\begin{equation}
  \label{a_09}
  \begin{split}
  a =   & \, \sqrt{ \dfrac{E^2 + C^4}{1 - C^2} -
        \left[ \dfrac{2 - C^2}{2(1 - C^2)} \right]^2  } \\
a_1 = & \, a - i \dfrac{2 - C^2}{2 (1 - C^2)}; \quad a_2 = -a_1^* \\%
b_i = & \, \dfrac{4 + a_i^2}{a_i^2 + 2 - a_i x_i}; \quad i = 1,2 \\ %
x_i = & \, \dfrac12 \left( a_i \pm \sqrt{4 + a_i^2} \right); \quad
i = 1,2
   \end{split}
  \end{equation}
Sign for $x_i$ in \eqref{a_09} is chosen so that $|x_i|<1$.


\section{}

When the hopping integral $C$ is small then amplitude $a_0(t)$
slowly exponentially decays. Below we get two approximate
expressions for the amplitude when $C^2 \ll 1$.

\begin{figure}
\begin{center}
  \includegraphics[width=70mm,angle=0]{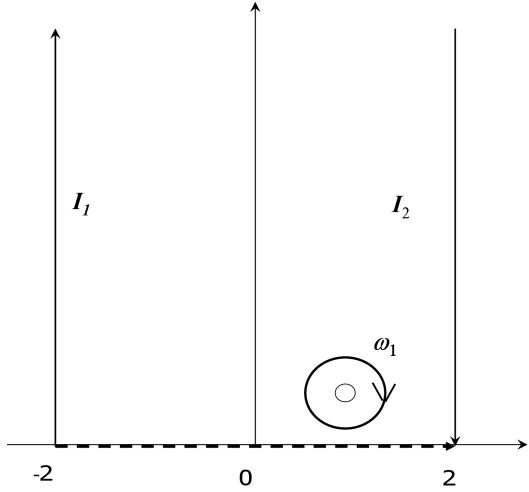}
 \caption{
The transformation of the initial integral \eqref{a_10} (dashed
line) into two integrals $I_1 \subset [-2; \, -2 + i\infty], \,
I_2 \subset [+2; \, +2+i\infty]$ and the residue in the pole
$\omega = \omega_1$.
        }
   \label{Fig_09_}
  \end{center}
\end{figure}

The first method consists in modification the integral formula
\eqref{a0_2} for $a_0(t)$. This expression can be rewritten as:
\begin{equation}
  \label{a_10}
  a_0(t) = \dfrac{C^2}{2\pi} \int\limits_{-2}^2 \, {\rm d} \omega
  \dfrac{\sqrt{4 - \omega^2} \, \exp(i \omega t)}
  {(\omega + E) (\omega + E - \omega C^2) + C^4} \,.
  \end{equation}
If $C$ is small then an approximate expression for $a_0(t)$ can be
written. To do this, the integration contour in \eqref{a_10}
should be deformed as shown in Fig.~\ref{Fig_09_}. In the upper
semiplane the integrand in \eqref{a_10} has a pole at the point
$\omega = \omega_1$:
\begin{equation}
  \label{om1}
  \omega_1 = \dfrac{1}{2 (1 - C^2)} \,
\left( -E (2 - C^2) + i \, C^2 \sqrt{4 - 4 C^2 - E^2} \right) \,.\end{equation}

When $C \ll 1$ then the pole lies very close to the real axis (on
the distance $\sim C^2$). Therefor the contribution of the pole
term decreases with time very slowly, -- slower then the
contribution from integrals $I_1$ and $I_2$ (see
Fig.~\ref{Fig_09_}).

The reason is that integrals $I_1$ and $I_2$ contain the
exponentially decaying multiplier $\exp(- \omega t)$. Therefor at
large time the contribution from these integrals is small in
comparison with the contribution from the pole. Thus, at small
$C^2$ and taking into account only the pole contribution, we get
the following approximate expression for the amplitude decay:
\begin{equation}
  \label{a_11}
  a_0(t) \approx \exp(i \omega t)\,
  \sqrt{\dfrac{4 - \omega_1^2}{4 - 4 \, C^2 - E^2}}.
  \end{equation}
At small $C^2$ the expression for the amplitude can be
additionally simplified and the result coincides with the Fermi's
golden rule:
\begin{equation}
  \label{a_12}
  a_0(t) \approx \exp
  \left[
   \left( - i E - \dfrac12 C^2 \sqrt{4 - E^2}  \right) t
  \right].
  \end{equation}

Other approximate expression for the amplitude $a_0(t)$ (at  $C^2
\ll 1$) can be derived using the perturbation theory for the
initial equation \eqref{a0_1}:
\begin{equation}
  \label{a_13}
  \dot a_0(t) = - i E a_0(t) - C^2 \int\limits_0^t
  B_0(t - t') \, a_0(t') \, {\rm d} t'; \quad B_0(t) = J_0(2t) +
  J_2(2t) \,.
  \end{equation}
Lets introduce the function $\widetilde a_0 = a_0 \exp(iEt)$. For
this function we have the following equation:
\begin{equation}
  \label{a_14}
  \dfrac{{\rm d} \widetilde a_0}{{\rm d}t} = -C^2 \int\limits_0^t
  B_0(t - t') \, a_0(t') \, {\rm d} t'.
  \end{equation}
Function $\widetilde a_0$ varies slowly at small $C$, and in the
first order of the perturbation theory we can put  $\widetilde
a_0(t') \approx  \widetilde a_0(t)$ in the integrand. As a result
the linear differential equation can be obtained. Its solution is:%
\begin{equation}
  \label{a_141}
 \widetilde a_0(t) = \exp
 \left[
 -C^2 \int\limits_0^t (t - t') \exp(i E t')
 [J_0(2t') + J_2(2t') ] \, {\rm d} t'
 \right].
  \end{equation}
Note that this expression is valid when time $t$ is small. If $t
\gg 1$ then expression \eqref{a_141} can be additionally
simplified. Evaluating  the corresponding integrals and returning
back to the function $a_0(t)$, we get:
\begin{equation}
  \label{a_15}
  a_0(t) = \exp
 \left[
 \dfrac{C^2}{2} - \dfrac{C^2 \sqrt{4 - E^2}}{2} \, t -
  i E \left( 1 + \dfrac{C^2}{2} \right) t +
  \dfrac{i E C^2}{2 \sqrt{4 - E^2}}
 \right].
  \end{equation}
Note that the error estimation of the exponent shows that this
error is of the order not exceeding $t^{-3/2}$.

\begin{figure}
\begin{center}
  \includegraphics[width=100mm,angle=0]{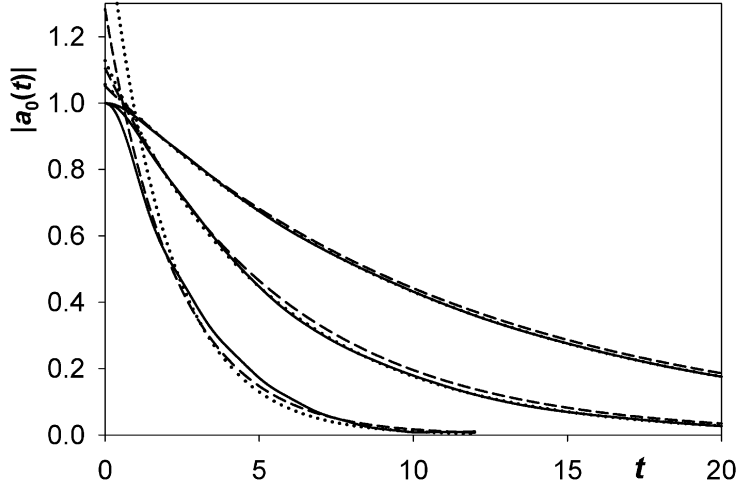}
 \caption{
The dependence of the amplitude $a_0(t)$ decay vs. time. Parameter
$E=1$. Parameter $C$ from up to down in the figure: $C^2=0.1; \,
0.2; \, 0.5$. Solid lines -- numerical results; dashed lines --
the pole approximations \eqref{a_11}; dotted lines -- the
perturbation theory \eqref{a_15}.
  }
  \label{Fig_10_}
  \end{center}
\end{figure}

It also should be pointed out that expression \eqref{a_15} for
$|a_0|^2$ gives an answer coinciding with the Fermi's golden rule
(but with extra multiplier $\exp(C^2)$ in  \eqref{a_15}).
Fig.~\ref{Fig_10_} shows the comparison of numerical answer for
the amplitude $a_0(t)$ decay with the approximate expressions
\eqref{a_11} and \eqref{a_15}.

\section{}

Consider now the case when $E > 2 -C^2, \,\, (C^2 < 2)$ and when
there exists one localized bounded state. We find the eigenstate
$\overrightarrow{\Psi}_{\rm loc}$ localized at the lattice end and
solve Schr\"odinger equation $H \overrightarrow{\Psi}_{\rm loc} =
\varepsilon \overrightarrow{\Psi}_{\rm loc}$ (for the hamiltonian
$H$ see \eqref{Ham1}). The solution has the form of exponentially
decaying function of $j$:
\begin{equation}
  \label{psi1}
  \overrightarrow{\Psi}_{\rm loc} = A \left\{ \dfrac{1}{C},
  \, \alpha^{j-1} \right\}, \quad
  A = \left( \dfrac{1}{C^2} + \dfrac{\alpha^2}{1 -
  \alpha^2} \right)^{-1/2}.
\end{equation}

Parameter $\alpha$ is expressed through the energy $\varepsilon$:
\begin{equation}
  \label{alf}
  \alpha = \dfrac12 \, \left( \varepsilon -
  \sqrt{\varepsilon^2 - 4} \right).
\end{equation}
Energy $\varepsilon$ is related to parameters $E$ and $C$ by the
relationship, coinciding with given by \eqref{a0_3}:
\begin{equation}
  \label{E}
  E = \varepsilon \, \left( 1 - \dfrac{C^2}{2} \right) +
  \dfrac{C^2}{2} \sqrt{\varepsilon^2 - 4}  \,.
\end{equation}

Expand now the total wave function in terms of the eigenfunctions
of the hamiltonian:
\begin{equation}
  \label{psi2}
  \overrightarrow{\Psi} = D \, \overrightarrow{\Psi}_{\rm loc} \,
  \exp(- i \varepsilon t) + {\rm C.C.S.},
\end{equation}
where C.C.S. means the contribution from the continuous spectrum.
The first term is just the contribution to amplitude from the
localized state. It is necessary to find the constant $D$. To do
this we consider the scalar product of the localized and the
complete wave functions. Because of the orthogonality of wave
functions one can get: $\left< \vec \Psi_{\rm loc} \vec \Psi
\right> = D \exp(- i \varepsilon t)$. In this equality we put
$t=0$. As the wave function is fully localized on the impurity
site at $t=0$, i.e. $\vec \Psi (t=0) = (1,0,0, \ldots)$, then it
follows that $D = \Psi_{\rm loc}^*(j=1)$. Thus, the necessary
contribution to the amplitude on the impurity site from the
localized state is:
\begin{equation}
  \label{a_16}
a_0^{\rm loc} = \left| \Psi_{\rm loc}(1) \right|^2 \exp(- i
\varepsilon
  t).
  \end{equation}
Substituting the expression $\Psi_{\rm loc}(1) = A C^{-1}$ from
\eqref{psi1} and expressing $A$ through $\varepsilon$
(see~\eqref{alf}) we get the final answer, coinciding with
\eqref{a0_3}:
\begin{equation}
  \label{a_17}
  a_0^{\rm loc}(t) = \dfrac{\exp(- i \varepsilon  t)}
  {1 + \dfrac{C^2}{2}
\left( \dfrac{\varepsilon}{\sqrt{\varepsilon^2 - 4}} -1 \right)}.
\end{equation}
%



\end{document}